\documentclass[aps,graphicx,twocolumn]{revtex4}
\usepackage{amsmath}
\usepackage{amscd}
\usepackage{graphicx}
\usepackage{multirow}
\usepackage{color}
\begin{document}

\title{Complete hyperentangled Bell state analysis for polarization and time-bin hyperentanglement}
\author{ Xi-Han Li$^{1,2}$\footnote{
Email address: xihanlicqu@gmail.com}, Shohini Ghose$^{2,3}$}
\address{$^1$ Department of Physics, Chongqing University,
Chongqing, China \\$^2$Department of Physics and Computer Science, Wilfrid Laurier University, Waterloo, Canada\\
$^3$ Institute for Quantum Computing, University of Waterloo, Canada}

\date{\today }
\begin{abstract}
We present a complete hyperentangled Bell state analysis protocol for two-photon four-qubit states which are simultaneously entangled in the polarization and time-bin degrees of freedom. The 16 hyperentangled states can be unambiguously distinguished via two steps. In the first step, the polarization entangled state is distinguished deterministically and nondestructively with the help of the cross-Kerr nonlinearity. Then, in the second step, the time-bin state is analyzed with the aid of the polarization entanglement. We also discuss the applications of our protocol for quantum information processing. Compared with hyperentanglement in polarization and spatial-mode degrees of freedom, the polarization and time-bin hyperentangled states provide saving in quantum resources since there is no requirement for two spatial modes for each photon. This is the first complete hyperentangled Bell state analysis scheme for polarization and time-bin hyperentangled states, and it can provide new avenues for high-capacity, long-distance quantum communication.
\end{abstract}
\maketitle

\section{Introduction}
Entanglement is a unique quantum phenomenon and a crucial resource widely used in quantum computation and quantum communication in the past decades. It plays a key role as the information carrier in quantum communication schemes such as quantum key distribution \cite{qkd1,qkd2}, quantum dense coding \cite{dense1,dense2}, quantum teleportation \cite{tele}, quantum secure direct communication \cite{qsdc1,qsdc2,qsdc3} and so on.
Among many physical systems proposed for quantum communication, the photon is the most competitive candidate due to its manipulability and high-speed transmission features. Photons have many different degrees of freedom (DOFs) to carry quantum information, such as, for instance, the polarization, time-bin, spatial-mode, frequency, and orbital angular monmentum. Entangled states have been prepared in each of these DOFs in experiments. Moreover, simultaneous entanglement in more than one of these DOFs, referred to as the hyperentanglement, has also been generated. In 1997, Kwiat \emph{et al}. proposed the first scheme to generate an energy-momentum-polarization hyperentangled state \cite{preparation1}. In 2005, Yang \emph{et al.} generated a two-photon state entangled both in polarization and spatial mode DOFs to realize the all-versus-nothing test of local realism \cite{preparation2}. In the same year, an experimental demonstration of a photonic hyperentangled system simultaneously entangled in polarization, spatial mode and time-energy was first reported \cite{preparation3}. Later, Vallone \emph{et al.} also realized a six-qubit hyperentangled state which was entangled in polarization and two longitudinal momentum DOFs \cite{preparation4}. The hyperentangled photons carry information encoded in more than one DOF at the same time and different DOFs can be manipulated independently. These distinct features make hyperentanglement useful for many applications in quantum information processing. It can increase the channel capacity and also enhance the security of quantum communication schemes \cite{ah1}. For instance, hyperentanglement can assist in the conventional Bell state analysis (BSA) by enlarging the Hilbert space \cite{CBSAH1,CBSAH2,CBSAH3,CBSAH4}. In 2008, it was exploited to beat the channel capacity limit in a protocol in which complete BSA of polarization states is aided by the orbital angular momentum \cite{CBSAH5}. Hyperentangled states have also been used to accomplish deterministic entanglement purification of polarization entanglement \cite{ah4,ah5,ah6,ah7}, construct hyperparallel photonic quantum computing \cite{ah2,ah3} and quantum repeaters \cite{ah8}.  Entanglement concentration and entanglement purification protocols for hyperentangled state have also been proposed with the aim of establishing maximally hyperentangled channels between distant parties \cite{hchp1,hchp2,hchp3,hchp4,hchp5,hchp6,hchp7,hchp8}.

In quantum communication schemes that are based on entanglement, state analysis is an indispensable step required to read out encrypted information. State analysis is of both theoretical significance and practical importance, and thus it has been the focus of much research. Although a set of mutually orthogonal states should in principal be deterministically distinguishable, this becomes a challenge in photonic systems since interaction between photons is not an easy task using current techniques. It was proved that complete BSA is impossible via linear optics alone \cite{bsa1,bsa2,bsa3}. Hyperentangled Bell state analysis (HBSA) in which states of two or more DOFs have to be distinguished simultaneously is even more difficult, and is thus also not possible via linear optics alone. It has been shown that 16 hyperentangled Bell states can be classed into only 7 groups with linear optics \cite{HBSA1,HBSA2}. Therefore, auxiliary states and assistant tools have to be utilized to accomplish complete state analysis \cite{BSAS,BSAN,CHBSA1,CHGSA,CHBSA2,CHBSA3}. In 2010, Sheng \emph{et al.} proposed the first complete HBSA scheme for polarization and spatial-mode hyperentangled states \cite{CHBSA1}. The two DOFs are distinguished with the help of the cross-Kerr nonlinearity in two steps. Later, an efficient hyperentangled Greenberger-Horne-Zeilinger (GHZ) state analysis scheme was presented \cite{CHGSA}. Moreover, complete HBSA can also be realized with the help of giant nonlinear optics in optical microcavities and nitrogen-vacancy centers in resonators \cite{CHBSA2,CHBSA3,CHBSA4}. Recently, Liu\emph{ et al.} proposed a complete nondestructive analysis assisted by cross-Kerr nonlinearity of   two-photon six-qubit hyperentangled Bell states in which the photons are entangled simultaneously in the polarization and two longitudinal momentum DOFs \cite{CHBSAT}.

So far, all hyperentangled state analysis protocols have dealt with the polarization and spatial-mode hyperentangled state. This kind of hyperentanglement is a promising candidate for quantum communication since both these two DOFs can be manipulated with high fidelity at present. However, if the spatial-mode DOF is exploited, each photon requires two paths during the transmission, which leads to a lot of extra requirements on resources in long-distance quantum communication. Instead, the time-bin DOF of photons with two different arrival times as the basis can save the extra resources. It is also a simple and conventional classical DOF of photons, and can be simply discriminated by the time of arrival. Despite the difficulties in manipulation of time-bin DOF, we have previously proposed a hyperentanglement concentration scheme for  polarization and time-bin hyperentanglement \cite{hchp8}.
Here, we propose the first complete HBSA scheme for polarization and time-bin hyperentangled states. The scheme consists of two steps. In the first step the polarization states are distinguished by two quantum nondemolition detectors (QNDs) constructed with the cross-Kerr nonlinearity. The parity and phase information of the polarization state are read without destroying the state. Then the time-bin state is analyzed with the help of the polarization entanglement, without resorting to any nonlinearity. The 16 hyperentangled Bell states can be completely and deterministically discriminated. We also give two examples of the application of our HBSA scheme for quantum information processing. Our protocol is single-shot and requires less nonlinearities compared with previous HBSA schemes for polarization and spatial-mode hyperentanglement, and is thus useful for practical long-distance quantum communication.

\section{Complete hyperentangled Bell state analysis for polarization and time-bin hyperentanglement}
The two-photon four-qubit hyperentangled Bell state can be written as
\begin{eqnarray}
  \vert \Upsilon \rangle_{AB} = \vert \Theta_P\rangle_{AB}\otimes \vert\Xi_T\rangle_{AB}.
\end{eqnarray}
Here $A$ and $B$ denote the two photons and the subscripts $P$ and $T$ represent the polarization and time-bin DOFs, respectively.
$\vert \Theta_P\rangle_{AB}$ is one of four Bell states in the polarization DOF,
\begin{eqnarray}
  \vert \Phi^{\pm}_P\rangle_{AB}&=&\frac{1}{\sqrt{2}} (\vert HH\rangle \pm \vert VV \rangle)_{AB}, \\
  \vert \Psi^{\pm}_P\rangle_{AB}&=&\frac{1}{\sqrt{2}} (\vert HV\rangle \pm \vert VH \rangle)_{AB}.
\end{eqnarray}
Here $\vert H\rangle$ and $\vert V\rangle$ indicate the horizontal and vertical polarizations, respectively. The time-bin state $\vert\Xi_S\rangle_{AB}$ is one of the four Bell states in the time-bin DOF,
\begin{eqnarray}
   \vert \Phi^{\pm}_T\rangle_{AB}&=&\frac{1}{\sqrt{2}} (\vert SS\rangle \pm \vert LL \rangle), \\
  \vert \Psi^{\pm}_T\rangle_{AB}&=&\frac{1}{\sqrt{2}} (\vert SL\rangle \pm \vert LS \rangle).
\end{eqnarray}
Here $\vert S\rangle$ and $\vert L\rangle$ denote the two different time-bins, the early ($S$) and the late ($L$). Taking the two DOFs together, there are 16 hyperentangled Bell states, which can be completely distinguished in the following two steps.

\subsection{Complete Bell state analysis for the polarization degree of freedom via cross-Kerr nonlinearity}

The principle of our proposed polarization BSA protocol is shown in Fig. 1. Two QNDs are used to read the parity and phase information of the polarization state. Each QND is composed of two polarizing beam splitters (PBSs), two nonlinearities and a coherent probe beam $\vert \alpha\rangle$.  The PBS transmits the horizontal state $\vert H\rangle$ and reflects the vertical one $\vert V\rangle$. Two photons are guided into two input ports labeled  $A$ and $B$, and then interact with the nonlinear medium after the first PBS. The interaction between the photons in the two paths and the coherent probe beam causes the coherent state to pick up a phase shift $\vert \alpha\rangle \rightarrow \vert \alpha e^{iN\theta}\rangle$ when there are $N$ photons in the corresponding spatial mode \cite{kerr}.
Here $\theta=\chi t$ where $\chi$ is the coupling strength of the nonlinearity and $t$ is the interaction time, both of which can be set in advance.
For example, if the polarization state is $\vert V \rangle_A\vert H\rangle_B$ ($\vert H \rangle_A\vert V\rangle_B$), both the two photons go through the upper (lower) path  and the coherent state $\alpha_1$ picks up a phase shift $2\theta$ (-$2\theta$). We can choose the $X$-quadrature measurement such that it cannot distinguish phase shifts differing only in sign ``$\pm$". This feature preserves the coherence of photons with respect to each other as well as the photons themselves. The evolutions of these four polarization Bell states in the first QND are
\begin{eqnarray}
  \vert \Phi^{\pm}_P\rangle \vert \alpha_1\rangle &=& \frac{1}{\sqrt{2}}(\vert HH\rangle \pm \vert VV \rangle)\vert \alpha_1\rangle\nonumber\\
  &\rightarrow& \frac{1}{\sqrt{2}}(\vert HH\rangle \pm \vert VV \rangle)\vert \alpha_1\rangle= \vert \Phi^{\pm}_P\rangle \vert \alpha_1\rangle, \\
  \vert \Psi^{\pm}_P\rangle \vert \alpha_1\rangle&=&\frac{1}{\sqrt{2}} (\vert HV\rangle \pm \vert VH \rangle)\vert \alpha_1\rangle\nonumber\\
   &\rightarrow& \frac{1}{\sqrt{2}} (\vert HV\rangle\vert \alpha_1e^{-2 i\theta}\rangle \pm \vert VH \rangle\vert \alpha_1e^{2 i\theta}\rangle)\nonumber\\
   &=& \vert \Psi^{\pm}_P\rangle \vert \alpha_1e^{\pm 2 i\theta}\rangle.
\end{eqnarray}

\begin{figure}[!h]
\centering
\includegraphics*[width=3.2in]{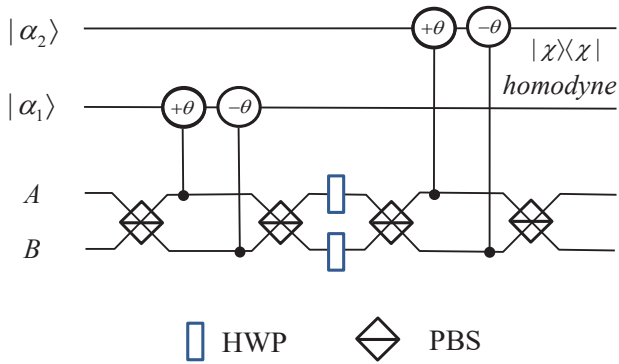}
\caption{Schematic diagram of the complete polarization Bell state analyzer. The polarizing beam splitters (PBSs) transmit horizontal polarized states while reflecting vertical ones. The half-wave plates (HWPs) implement the Hadamard operation, which transform the phase information of the state into the parity information. The four cross-Kerr nonlinear interactions put phase shift $\pm \theta$ on the coherent states $\vert \alpha_1 \rangle$ and $\vert \alpha_2 \rangle$ if a photon appears in the corresponding spatial modes. The first QND is used to distinguish $\vert \Phi^{\pm}_P\rangle$ from $\vert \Psi^{\pm}_P\rangle$ while the second QND reads the relative phase information ``$\pm$". After the homodyne measurements on the two coherent states, the four polarization Bell states can be completely distinguished without destroying the entanglement or losing the photons.}
\end{figure}

By measuring the phase shift via the $X$-quadrature measurement, the parity information of the polarization state can be distinguished, i.e., $\vert \Phi^{\pm}_P\rangle$ is distinguished from $\vert \Psi^{\pm}_P\rangle$. Another PBS changes the state back to it's original spatial status, i.e., one photon per path. The photons and polarization states are preserved after the first QND. Then two half-wave plates (HWPs) implement the Hadamard operation
\begin{eqnarray}
  \vert H\rangle &\rightarrow& \frac{1}{\sqrt{2}}(\vert H\rangle+\vert V\rangle), \\
  \vert V\rangle &\rightarrow& \frac{1}{\sqrt{2}}(\vert H\rangle-\vert V\rangle).
\end{eqnarray}
The polarization state changes as follows: both $\vert \Phi^+_P\rangle$ and $\vert \Psi^-_P\rangle$ are invariant while $\vert \Phi^-_P\rangle\rightleftharpoons\vert \Psi^+_P\rangle$. In other words, the two HWPs transfer the phase information of the state to the parity one. After the second QND, the phase information is obtained. That is, the original $\vert \Phi^+_P\rangle$ and $\vert \Psi^+_P\rangle$ states will lead to no phase shift on $\vert \alpha_2\rangle$ while the other two states put $\pm 2\theta$ on the coherent state. Hence, the four polarization Bell states are unambiguously discriminated. The relation between the original state, the two phase shifts of the two coherent beams, and the new state are shown in Table. I. The preserved polarization entanglement will play an important role in the time-bin state analysis.
\begin{table}[!h]
\caption{Relations between the original state, the new state of the polarization DOF, and the two phase shifts of the coherent states. }
\begin{tabular}{c|ccccc|ccccc|c}
\hline\hline
Original state & & & $\vert \alpha_1\rangle$ & & & & &  $\vert \alpha_2\rangle$ & & & New state\\
\hline
$\vert \Phi^{+}_P\rangle$ & & & 0 & & & & &  0& & &$\vert \Phi^{+}_P\rangle$\\

$\vert \Phi^{-}_P\rangle$ & & & 0 & & & & &  $\pm2\theta$& & & $\vert \Psi^{+}_P\rangle$\\

$\vert \Psi^{+}_P\rangle$ & & & $\pm2\theta$ &  & & & & 0&  & &$\vert \Phi^{-}_P\rangle$\\

$\vert \Psi^{-}_P\rangle$ & & & $\pm2\theta$ & & & & & $\pm2\theta$&  & &$\vert \Psi^{-}_P\rangle$\\
\hline
\end{tabular}
\end{table}

\subsection{Complete Bell state analysis for the time-bin state assisted by the polarization entanglement }
In the second step, the two photons are each projected onto single-photon Bell basis states and the time-bin entangled states are distinguished with the help of the information about the polarization entanglement.

Each photon has two DOFs and can be measured in an entangled basis composed of the two DOFs,
\begin{eqnarray}
  \vert \phi^{\pm}\rangle_X &=& \frac{1}{\sqrt{2}}(\vert HL\rangle \pm\vert VS\rangle)_X, \\
  \vert \psi^{\pm}\rangle_X &=& \frac{1}{\sqrt{2}}(\vert HS\rangle \pm\vert VL\rangle)_X.
\end{eqnarray}
Here $X$ can be either $A$ or $B$. The single-photon Bell states can be completely distinguished by a single-photon Bell state analyser (SPBSA), shown in Fig. 2.

\begin{figure}[!h]
\centering
\includegraphics*[width=3.2in]{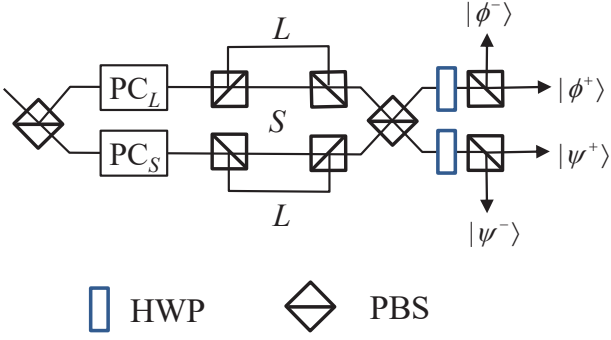}
\caption{Schematic diagram of the single-photon Bell state analyzer (SPBSA). PC$_L$ (PC$_S$) is a Pockel cell which effects a bit flip operation when the $L(S)$ component is present. Then two unbalanced interferometers composed of two PBSs are used to adjust the time-bin states: the length difference between the long $(L)$ path and the short $S$ one is set to cancel the time interval between two time-bins. Then the two paths intersect at a PBS and the photon is measured in the diagonal polarization basis.}
\end{figure}

Two Pockel cells (PCs) \cite{PC} are used to flip the polarizations of the photons at specific times, i.e., the PC$_L$ (PC$_S)$ is activated only when the $L$ $(S)$ component is present. Then two unbalanced interferometers composed of two PBSs are used to adjust the time-bin state such that the path length difference between the long path $L$ and the short one $S$ cancels the time difference between the two time-bins. The HWPs effect the Hadamard operation, and the PBSs then allow a measurement in the diagonal basis $\vert \pm\rangle=\frac{1}{\sqrt{2}}(\vert H\rangle \pm\vert V\rangle)$. As shown in Fig. 2, different single-photon Bell states will trigger different detectors placed on the four output ports. Thus the four single-photon Bell states can be deterministically distinguished.

Using two SPBSAs for each of the two photons, the time-bin state can be discriminated with the help of the undestroyed polarization entanglement. Each of the potential hyperentangled Bell states will result in four possible detections. There are 16 possible measurement combinations, which can be classed into four groups. Each group corresponds to four specific hyperentangled Bell states. The detailed relations are shown in Table. II. With the knowledge of the polarization entanglement, the time-bin state can be deterministically identified. For example, if the two measurement results are $\vert \psi^-\rangle_A$ and $\vert \phi^+\rangle_B$, the new state after the first step belongs to the last group. If the first step determines that the original polarization state is $\vert \Phi^-_P\rangle$ (the new state is $\vert \Psi^+_P\rangle$ ), one can deduce that the time-bin state is $\vert \Phi^-_T\rangle$. The initial hyperentangled state is $\vert \Phi^-_P\rangle \otimes\vert \Phi^-_T\rangle$.

 \begin{table}[!h]
\caption{Relations between the new state before the second step and possible detections. }
\begin{tabular}{cc|cc}
\hline\hline
New states & & & Possible detections  \\
\hline
$\vert \Phi^{+}_P\rangle\otimes \vert \Phi^{+}_T\rangle$, $\vert \Phi^{-}_P\rangle\otimes \vert \Phi^{-}_T\rangle$,& & &
$\vert \phi^{+}\rangle_A\vert \phi^{+}\rangle_B$, $\vert \phi^{-}\rangle_A\vert \phi^{-}\rangle_B$, \\
$\vert \Psi^{+}_P\rangle\otimes \vert \Psi^{+}_T\rangle$,$\vert \Psi^{-}_P\rangle\otimes \vert \Psi^{-}_T\rangle$. & & &
$\vert \psi^{+}\rangle_A\vert \psi^{+}\rangle_B$, $\vert \psi^{-}\rangle_A\vert \psi^{-}\rangle_B$. \\
\hline
$\vert \Phi^{+}_P\rangle\otimes \vert \Psi^{+}_T\rangle$, $\vert \Phi^{-}_P\rangle\otimes \vert \Psi^{-}_T\rangle$, & & &
 $\vert \phi^{+}\rangle_A\vert \psi^{+}\rangle_B$, $\vert \phi^{-}\rangle_A\vert \psi^{-}\rangle_B$, \\
 $\vert \Psi^{+}_P\rangle\otimes \vert \Phi^{+}_T\rangle$, $\vert \Psi^{-}_P\rangle\otimes \vert \Phi^{-}_T\rangle$. & & &
 $\vert \psi^{+}\rangle_A\vert \phi^{+}\rangle_B$, $\vert \psi^{-}\rangle_A\vert \phi^{-}\rangle_B$. \\
\hline
$\vert \Phi^{+}_P\rangle\otimes \vert \Phi^{-}_T\rangle$, $\vert \Phi^{-}_P\rangle\otimes \vert \Phi^{+}_T\rangle$,& & &
 $\vert \phi^{+}\rangle_A\vert \phi^{-}\rangle_B$, $\vert \phi^{-}\rangle_A\vert \phi^{+}\rangle_B$,\\
 $\vert \Psi^{+}_P\rangle\otimes \vert \Psi^{-}_T\rangle$, $\vert \Psi^{-}_P\rangle\otimes \vert \Psi^{+}_T\rangle$. & & &
 $\vert \psi^{+}\rangle_A\vert \psi^{-}\rangle_B$, $\vert \psi^{-}\rangle_A\vert \psi^{+}\rangle_B$. \\
\hline
$\vert \Phi^{+}_P\rangle\otimes \vert \Psi^{-}_T\rangle$, $\vert \Phi^{-}_P\rangle\otimes \vert \Psi^{+}_T\rangle$,& & &
 $\vert \phi^{+}\rangle_A\vert \psi^{-}\rangle_B$, $\vert \phi^{-}\rangle_A\vert \psi^{+}\rangle_B$,\\
 $\vert \Psi^{+}_P\rangle\otimes \vert \Phi^{-}_T\rangle$, $\vert \Psi^{-}_P\rangle\otimes \vert \Phi^{+}_T\rangle$. & & &
  $\vert \psi^{+}\rangle_A\vert \phi^{-}\rangle_B$, $\vert \psi^{-}\rangle_A\vert \phi^{+}\rangle_B$. \\
\hline
\end{tabular}
\end{table}

From the preceding analysis, the 16 hyperentangled Bell states are unambiguously discriminated with our two-step scheme. The distinguishing of polarization state resorts to the cross-Kerr nonlinearity while the discrimination of time-bin state can be accomplished with the help of the preserved polarization entanglement, without any nonlinear optics.

\section{Applications of our complete HBSA scheme}
Hyperentangled states have a lot of applications in quantum communication in which a HBSA may required to read out the information. Here we demonstrate the applications of our HBSA for two protocols: quantum teleportation and entanglement swapping. Conventional quantum teleportation can transmit an unknown quantum state through a pre-established entangled channel without transmitting the photon itself.  Entanglement swapping is an important constituent for quantum repeaters in long-distance communication. The channel capacity of both protocols can be increased with hyperentangled states. Here we discuss the applications of our HBSA scheme in quantum teleportation and entanglement swapping using hyperentanglement in polarization and time-bin DOFs.

\subsection{Teleportation of a single-photon two-qubit state}
Suppose the sender Alice and the receiver Bob share a hyperentangled channel in advance
\begin{eqnarray}
  \vert \Upsilon\rangle_{AB} &=& \frac{1}{\sqrt{2}}(\vert HH\rangle+\vert VV\rangle)\otimes \frac{1}{\sqrt{2}}(\vert SS\rangle+\vert LL\rangle).
\end{eqnarray}
The unknown state of photon $X$ which Alice wants to send to Bob is
\begin{eqnarray}
\vert \varphi\rangle_X=(\alpha\vert H\rangle+\beta\vert V\rangle)\otimes(\delta\vert S\rangle+\eta\vert L\rangle).
\end{eqnarray}

Alice can either send the photon directly through a noisy channel or take advantage of the shared hyperentanglement. If she chooses the latter, she performs the HBSA on photons $X$ and $A$.  The state of the whole system can be rewritten  in terms of the hyperentangled state as
\begin{eqnarray}
 &&\vert \varphi\rangle_X \otimes \vert \Upsilon\rangle_{AB} \nonumber\\ &=& \frac{1}{4}[\vert \Phi^+_P\rangle_{XA}(\alpha \vert H\rangle+\beta\vert V\rangle)_B+\vert \Phi^-_P\rangle_{XA}(\alpha \vert H\rangle-\beta\vert V\rangle)_B\nonumber\\ &&+\vert \Psi^+_P\rangle_{XA}(\alpha \vert V\rangle+\beta\vert H\rangle)_B+\vert \Psi^-_P\rangle_{XA}(\alpha \vert V\rangle-\beta\vert H\rangle)_B ]\nonumber\\&& \otimes[\vert \Phi^+_T\rangle_{XA}(\delta \vert S\rangle+\eta\vert L\rangle)_B+\vert \Phi^-_T\rangle_{XA}(\delta \vert S\rangle-\eta\vert L\rangle)_B\nonumber\\ &&+\vert \Psi^+_T\rangle_{XA}(\delta \vert L\rangle+\eta\vert S\rangle)_B+\vert \Psi^-_T\rangle_{XA}(\delta \vert L\rangle-\eta\vert S\rangle)_B ].\nonumber\\
\end{eqnarray}

Alice has 16 possible measurement results, corresponding to which there are 16 potential single-photon two-qubit states for Bob's photon. With our HBSA scheme, the 16 hyperenatngled states can be completely distinguished, according to which Bob knows the state of his own photon with certainty. For example, if Alice's measurement result is $\vert \Psi^+_P\rangle_{XA}\otimes \vert \Phi^+_T\rangle_{XA}$, Bob's state is $(\alpha\vert V\rangle+\beta\vert H\rangle)_B\otimes (\delta\vert S\rangle+\eta\vert L\rangle)_B$. Then the original state can be obtained by Bob with proper single-photon unitary operations on both DOFs.

Compared with conventional quantum teleportation, the use of polarization and time-bin hyperentanglement provides a way for teleporting a single-photon two-qubit state which carries more information. In conventional quantum teleportation, two of four Bell states can be distinguished by linear optics alone. Thus Bob can get the desired state with 50$\%$ probability. In our sceme, without a complete HBSA, the 16 hyperentangled states can only be classed into 7 groups and each of them contains more than one state. Thus the uncertainty of Alice's measurement outcomes results in a mixed state in Bob's hand. This shows that a complete HBSA is indispensable for teleporting a single-photon two-qubit state.

\subsection{Entanglement swapping between hyperentangled pairs}
Entanglement swapping enables two independent photons to be entangled with each other without any direct interactions between them. This plays a crucial role in quantum repeaters. Conventional quantum repeaters establish entanglement in one DOF between two distant parties. Since hyperentanglement shared between two remote parties can improve the channel capacity greatly, it is useful to establish hyperentanglement between remote parties. Suppose two remote parties Alice and Bob each share hyperentanglement with a central node Charlie.
\begin{eqnarray}
  \vert \Upsilon\rangle_{AC_1} &=& \frac{1}{\sqrt{2}}(\vert HH\rangle+\vert VV\rangle)_{AC_1}\otimes \frac{1}{\sqrt{2}}(\vert SS\rangle+\vert LL\rangle)_{AC_1} , \nonumber
\\ \\
  \vert \Upsilon\rangle_{C_2B} &=& \frac{1}{\sqrt{2}}(\vert HH\rangle+\vert VV\rangle)_{C_2B}\otimes \frac{1}{\sqrt{2}}(\vert SS\rangle+\vert LL\rangle)_{C_2B}. \nonumber \\
\end{eqnarray}
Here $A$ and $B$ belong to Alice and Bob, respectively, and $C_1$, $C_2$ are held by Charlie. After Charlie performs HBSA on $C_1$ and $C_2$, Alice and Bob's photons will collapsed into a hyperentangled state.
\begin{eqnarray}
  &&\vert \Upsilon\rangle_{AC_1} \otimes \vert \Upsilon\rangle_{C_2B} \nonumber\\
  &=&\frac{1}{4}[(\vert \Phi^+_P\rangle_{C_1C_2}\otimes\vert \Phi^+_P\rangle_{AB}+\vert \Phi^-_P\rangle_{C_1C_2}\otimes\vert \Phi^-_P\rangle_{AB}\nonumber\\
  &&+ \vert \Psi^+_P\rangle_{C_1C_2}\otimes\vert \Psi^+_P\rangle_{AB}+\vert \Psi^-_P\rangle_{C_1C_2}\otimes\vert \Psi^-_P\rangle_{AB})\nonumber\\
  && \otimes(\vert \Phi^+_T\rangle_{C_1C_2}\otimes\vert \Phi^+_T\rangle_{AB}+\vert \Phi^-_T\rangle_{C_1C_2}\otimes\vert \Phi^-_T\rangle_{AB}\nonumber\\
  && +\vert \Psi^+_T\rangle_{C_1C_2}\otimes\vert \Psi^+_T\rangle_{AB}+\vert \Psi^-_T\rangle_{C_1C_2}\otimes\vert \Psi^-_T\rangle_{AB})].
\end{eqnarray}
From the expression we find the state of $AB$ depends on the measurement result of $C_1C_2$. For instance, if Charlie's result is $\vert \Phi^+_P\rangle_{C_1C_2}\otimes \vert \Psi^+_T\rangle_{C_1C_2}$, the state shared by Alice and Bob is $\vert \Phi^+_P\rangle_{AB}\otimes \vert \Psi^+_T\rangle_{AB}$. With Charlie's information, Alice and Bob can share the desired hyperentangled state with or without some additional single-photon operations. With our HBSA scheme, hyperentanglement can be established between distant parties, which will be useful for long-distance, high-capacity quantum communication.

\section{discussion and summary}
In this paper, we have proposed a HBSA protocol for polarization and time-bin hyperentanglement, in which the hyperentangled states  are distinguished in two steps. The polarization state is analyzed with the help of the cross-Kerr nonlinearity. Both the photons and the polarization entanglement are preserved after the discrimination. Then the time-bin state is distinguished with the help of the information about the polarization entanglement. No nonlinear interaction is required to analyze the time-bin entanglement. The two photons are each measured in the single-photon Bell basis. Based on this, the 16 hyperentangled states can be discriminated unambiguously.

The cross-Kerr nonlinearity is used in our scheme in the first step.  The feasibility of our HBSA scheme mainly depends on the cross-Kerr nonlinearity. Although it remains a challenge with current technology, it has been exploited in many quantum information processing protocols \cite{kerr,ah4,hchp7,CHBSA1,CHGSA}. Recent studies also show promising results for using the effect in the near future \cite{kerr1,kerr1+,kerr2,kerr3,kerr4,kerr5,kerr6}. Moreover, we note that in our HBSA scheme, only a small nonlinearity is required, as long as the phase shift can be distinguished from the zero phase shift case. This makes it more promising to implement with present techniques.

It is interesting to compare our HBSA with previous protocols that use different DOFs. First, in all the HBSA schemes the two DOFs are distinguished in two steps. It is important to point out that our scheme is one-shot in that no pause is required between two steps. In contrast, the other HBSA schemes need to confirm the spatial mode of the photons before they move on to the polarization state analysis \cite{CHBSA1,CHBSA2}. Otherwise, setups with more nonlinearities need to be prepared in advance for all possible situations. Second, in our scheme the discrimination of the second DOF is realized without any nonlinear optics while in other schemes the analysis of both DOFs resorts to nonlinearities \cite{CHBSA1,CHBSA2}. These two advantages make our scheme time-saving and resource-saving and thus  more useful and practical.

In summary, we have described an efficient scheme for complete HBSA of a two-photon system hyperentangled in the polarization and time-bin DOFs. The 16 hyperentangled states can be distinguished unambiguously.
Our HBSA protocol is the first one to deal with polarization and time-bin hyperentanglement. Since the time-bin DOF requires less resource in long-distance quantum communication compared with the popular spatial-mode DOF, we believe it will find more applications in quantum communication schemes, thus making our HBSA protocol useful and relevant for future applications.

\section*{Acknowledgement}

XL is supported by the National Natural Science
Foundation of China under Grant Nos. 11574038 and 11547305. SG acknowledges support from he Natural Sciences and Engineering Research Council of Canada.


\begin{thebibliography}{99}
\bibitem{qkd1} A. K. Ekert, Phys. Rev. Lett. 67, 661 (1991).
\bibitem{qkd2} C. H. Bennett, G. Brassard, and N. D. Mermin, Phys. Rev. Lett. 68, 557 (1992).
\bibitem{dense1} C. H. Bennett and S. J. Wiesner, Phys. Rev. Lett. 69, 2881 (1992).
\bibitem{dense2} X. S. Liu, G. L. Long, D. M. Tong, and L. Feng, Phys. Rev. A 65, 022304 (2002).
\bibitem{tele} C. H. Bennett, G. Brassard, C. Crepeau, R. Jozsa, A. Peres, and W. K. Wootters, Phys. Rev. Lett. 70, 1895 (1993).
\bibitem{qss1} M. Hillery, V. Bu\v{z}ek, and A. Berthiaume, Phys. Rev. A 59, 1829 (1999).
\bibitem{qss2} A. Karlsson, M. Koashi, and N. Imoto, Phys. Rev. A 59, 162 (1999).
\bibitem{qss3} L. Xiao, G. L. Long, F. G. Deng, and J. W. Pan, Phys. Rev. A 69, 052307 (2004).
\bibitem{qsdc1} G. L. Long and X. S. Liu, Phys. Rev. A 65, 032302 (2002).
\bibitem{qsdc2} F. G. Deng, G. L. Long, and X. S. Liu, Phys. Rev. A 68, 042317 (2003).
\bibitem{qsdc3} C. Wang, F. G. Deng, Y. S. Li, X. S. Liu, and G. L. Long, Phys. Rev. A 71, 044305 (2005).
\bibitem{preparation1} P. G. Kwiat, J. Mod. Opt. 44, 2173 (1997).
\bibitem{preparation2} T. Yang, Q. Zhang, J. Zhang, J. Yin, Z. Zhao, M. $\dot{Z}$ukowski, Phys. Rev. Lett. 95, 240406 (2005).
\bibitem{preparation3} J. T. Barreiro, N. K. Langford, N. A. Peters, and P. G. Kwiat, Phys. Rev. Lett. 95, 260501 (2005).
\bibitem{preparation4} G. Vallone, R. Ceccarelli, F. De Martini, and P. Mataloni, Phys. Rev. A 79, 030301R (2009).
\bibitem{ah1} S. P. Walborn, M. P. Almeida, P. H. Souto Ribeiro, and C. H. Monken, Quantum Inf. Comput. 6, 336 (2006).

\bibitem{CBSAH1} P. G. Kwiat and H. Weinfurter, Phys. Rev. A 58, R2623 (1998).
\bibitem{CBSAH2} S. P. Walborn, S. P\'{a}dua, and C. H. Monken, Phys. Rev. A 68, 042313 (2003).
\bibitem{CBSAH3} C. Schuck, G. Huber, C. Kurtsiefer, and H. Weinfurter, Phys. Rev. Lett. 96, 190501 (2006).
\bibitem{CBSAH4} M. Barbieri, G. Vallone, P. Mataloni, and F. De Martini, Phys. Rev. A 75, 042317 (2007).
\bibitem{CBSAH5} J. T. Barreiro, T. C. Wei, and P. G. Kwiat, Nat. Phys. 4, 282 (2008).
\bibitem{ah4} Y. B. Sheng and F. G. Deng, Phys. Rev. A 81, 032307 (2010).
\bibitem{ah5} X. H. Li, Phys. Rev. A 82, 044304 (2010).                                         ah4,hchp7,CHBSA1,CHGSA,Kerr
\bibitem{ah6} Y. B. Sheng and F. G. Deng, Phys. Rev. A 82, 044305 (2010).
\bibitem{ah7} F. G. Deng, Phys. Rev. A 83, 062316 (2011).
\bibitem{ah2} B. C. Ren and F. G. Deng, Sci. Rep. 4, 4623 (2014).
\bibitem{ah3} B. C. Ren, H. R. Wei, and F. G. Deng, Laser Phys. Lett. 10, 095202 (2013).
\bibitem{ah8} T. J. Wang, S. Y. Song, G. L. Long, Phys. Rev. A 85, 062311 (2012).
\bibitem{hchp1} B. C. Ren, F. F. Du, F. G. Deng, Phys. Rev. A 90, 052309 (2014).
\bibitem{hchp2} B. C. Ren, F. G. Deng, Laser Phys. Lett. 10, 115201 (2013).
\bibitem{hchp3} B. C. Ren, G. L. Long, Opt. Express 22, 6547 (2014).
\bibitem{hchp4} T. J. Wang, C. Cao, C. Wang, Phys. Rev. A 89, 052303 (2014).
\bibitem{hchp5} B. C. Ren, F. F. Du, F. G. Deng, Phys. Rev. A 88, 012302 (2013).
\bibitem{hchp6} X. H. Li, S. Ghose, Laser Phys. Lett. 11, 125201 (2014).
\bibitem{hchp7} X. H. Li, S. Ghose, Opt. Express 23, 3550 (2014).
\bibitem{hchp8} X. H. Li, S. Ghose, Phys. Rev. A 91, 062302 (2015).
\bibitem{bsa1} N. L\"{u}tkenhaus, J. Calsamiglia, and K. A. Suominen, Phys. Rev. A 59, 3295 (1999).
\bibitem{bsa2} J. Calsamiglia, Phys. Rev. A 65, 030301(R) (2002).
\bibitem{bsa3} K. Mattle, H. Weinfurter, P. G. Kwiat, and A. Zeilinger, Phys. Rev. Lett. 76, 4656 (1996).
\bibitem{HBSA1} T. C. Wei, J. T. Barreiro, P. G. Kwiat, Phys. Rev. A 75(R), 060305 (2007).
\bibitem{HBSA2} N. Pisenti, C. P. E. Gaebler, T. W. Lynn, Phys. Rev. A 84, 022340 (2011).
\bibitem{BSAS} F. Ewert and P. van Loock, Phys. Rev. Lett. 113, 140403 (2014).
\bibitem{BSAN} S. D. Barrett, P. Kok, K. Nemoto, R. G. Beausoleil, W. J. Munro, and T. P. Spiller, Phys. Rev. A 71, 060302(R) (2005).
\bibitem{CHBSA1} Y. B. Sheng, F. G. Deng, and G. L. Long, Phys. Rev. A 82, 032318 (2010).
\bibitem{CHGSA} Y. Xia, Q. Q. Chen, J. Song, H. S. Song, J. Opt. Soc. Am. B 29, 1029–1037 (2012).
\bibitem{CHBSA2} B. C. Ren, H. R. Wei, M. Hua, T. Li, and F. G. Deng, Opt. Express 20, 24664 (2012).
\bibitem{CHBSA3} T. J. Wang, Y. Lu, G. L. Long, Phys. Rev. A 86, 042337 (2012).
\bibitem{CHBSA4} Q. Liu and M. Zhang, Phys. Rev. A 91, 062321 (2015).
\bibitem{CHBSAT} Q. Liu, G.Y. Wang, M. Zhang, F.G. Deng, arXiv:1511.00094
\bibitem{kerr} K. Nemoto and W. J. Munro, Phys. Rev. Lett. 93, 250502 (2004).
\bibitem{PC} D. Kalamidas, Phys. Lett. A 343, 331 (2005).
\bibitem{kerr1} H. F. Hofmann, K. Kojima, S. Takeuchi, and K. Sasaki, J. Opt. B 5, 218 (2003).
\bibitem{kerr1+} C. Wittmann, U. L. Andersen, M. Takeoka, D. Sych, and G. Leuchs, Phys. Rev. A 81, 062338 (2010).
\bibitem{kerr2} B. He, Q. Lin, and C. Simon, Phys. Rev. A 83, 053826 (2011).
\bibitem{kerr3} A. Feizpour, X. Xing, and A. M. Steinberg, Phys. Rev. Lett. 107, 133603 (2011).
\bibitem{kerr4} C. Zhu and G. Huang, Opt. Express 19, 23364 (2011).
\bibitem{kerr5} Hoi, I. C. et al. Phys. Rev. Lett. 111, 053601 (2013). 
\bibitem{kerr6} Sathyamoorthy, S. R.et al. Phys. Rev. Lett. 112, 093601 (2014). 


\end{thebibliography}
\end{document}